\documentclass[12pt]{article}
\usepackage{graphics}
\usepackage{cite,a4wide}
\newcommand{\beq}{\begin{equation}}
\newcommand{\eeq}{\end{equation}}
\newcommand{\bea}{\begin{eqnarray}}
\newcommand{\eea}{\end{eqnarray}}
\renewcommand{\d}{\delta}

\renewcommand{\b}{\beta}
\renewcommand{\a}{\alpha}
\renewcommand{\ni}{\noindent}

\newcommand{\m}{\mu}

\newcommand{\q}{{\pi\over 5}}
\newcommand{\s}{\sigma}
\newcommand{\D}{\Delta}

\newcommand{\tU}{\tilde{U}}

\newcommand{\oh}{\frac{1}{2}}

\newcommand{\dg}{\dagger}
\newcommand{\non}{\nonumber}

\newcommand{\rf}[1]{(\ref{#1})}
\newcommand{\ra}{\rightarrow}

\begin{document}

\hfill October 1999

\begin{center}

\vspace{32pt}

  { \bf \Large The Vortex-Finding Property of Maximal \\
\bigskip
               Center (and Other) Gauges }

\end{center}

\vspace{18pt}

\begin{center}
{\sl M. Faber${}^a$, J. Greensite${}^{bc}$, {\v S}. Olejn\'{\i}k${}^d$,
and D. Yamada${}^{b}$}
\end{center}

\vspace{18pt}

\begin{tabbing}

{}~~~~~~~~~~~~~~~\= blah  \kill
\> ${}^a$ Inst. f\"ur Kernphysik, Technische Universit\"at Wien, \\
\> ~~A-1040 Vienna, Austria.  E-mail: {\tt faber@kph.tuwien.ac.at} \\
\\
\> ${}^b$ Physics and Astronomy Department, San Francisco \\
\> ~~State University, San Francisco, CA 94117 USA.  \\
\> ~~E-mail: {\tt greensit@quark.sfsu.edu, dyamada@quark.sfsu.edu} \\
\\
\> ${}^c$ Theory Group, Lawrence Berkeley National Laboratory, \\
\> ~~Berkeley, CA 94720 USA.  E-mail: {\tt greensit@thsrv.lbl.gov} \\
\\
\> ${}^d$ Institute of Physics, Slovak Academy of Sciences, \\
\> ~~SK-842 28 Bratislava, Slovakia.  E-mail: {\tt fyziolej@savba.sk} 

\end{tabbing}

\vspace{18pt}

\begin{center}

{\bf Abstract}

\end{center}

\vspace{24pt}

   We argue that the ``vortex-finding'' property of maximal center
gauge, i.e.\ the ability of this gauge to locate center vortices
inserted by hand on any given lattice, is the key to its success in
extracting the vortex content of thermalized lattice configurations.
We explain how this property comes about, and why it is expected not
only in maximal center gauge, but also in an infinite class of gauge
conditions based on adjoint-representation link variables.  In
principle, the vortex-finding property can be foiled by Gribov copies.
This fact is relevant to a gauge-fixing procedure devised by
Kov\'{a}cs and Tomboulis, where we show that the loss of center
dominance, found in their procedure, is explained by a corresponding
loss of the vortex-finding property.  The association of center
dominance with the vortex-finding property is demonstrated numerically
in a number of other gauges.

\vfill

\newpage

\section{Introduction}

   Numerical evidence in favor of the center vortex theory of 
confinement has been steadily accumulating over the past three years 
\cite{indirect,Zako,Jan98,mog,Cas,dFE,Lang1,Tubby,bertle,ITEP,ET}. 
Underlying most of these numerical studies is a technique 
for locating center vortices in thermalized lattice configurations, 
known as center projection in maximal center gauge.  

   In its ``direct'' version \cite{Zako,Jan98}, maximal center gauge 
is the gauge in which 
\beq
       R = \sum_x \sum_\m \Bigl| \mbox{Tr}[U_\m(x)] \Bigr|^2
             ~~~ \mbox{is a maximum}
\label{e1}
\eeq
This gauge brings each link variable as close as possible, on average, to
a $Z_N$ center element, while preserving a residual $Z_N$ gauge invariance.
Center projection is a mapping of each SU($N$) link variable to
the closest $Z_N$ center element; e.g.\ in SU(2) gauge theory,
center projection is the mapping
\beq
        U_\m(x) \ra Z_\m(x) \equiv \mbox{signTr}[U_\m(x)]
\label{e2}
\eeq
The excitations on the projected $Z_N$ lattice are point-like, 
line-like, or surface-like objects, in $D=2,3$, or $4$ dimensions 
respectively, known as ``P-vortices.''
These are thin objects, only one lattice spacing across.
There is substantial numerical evidence, for SU(2) gauge theory, that thin
P-vortices lie roughly in the middle of thick center vortices on the 
unprojected lattice, and that these thick vortices produce the entire 
asymptotic SU(2) string tension \cite{Jan98}. The number of
P-vortices mod 2 linking a large loop is closely correlated with the 
sign of the corresponding SU(2) Wilson loop
\cite{Jan98,dFE}, and P-vortices themselves are regions of high action on 
the unprojected lattice \cite{mog}.  It is found that removal of
center vortices not only removes the asymptotic string tension, but chiral 
symmetry goes as well, and the SU(2) lattice is then brought to trivial 
topology \cite{dFE}.  
The vortex density has been found to scale as predicted by asymptotic
freedom \cite{Lang1},\cite{Jan98,mog}, and, at finite temperature, 
the non-vanishing
string tension of spatial Wilson loops in the deconfined phase can be 
understood in terms of vortices winding through the periodic time direction
\cite{Tubby,bertle}.  Vortex percolation properties at finite temperature
have also been studied in ref.\ \cite{ITEP}.  The world-lines of
abelian-projection monopoles are found to lie on P-vortex surfaces,
and the field-strength associated with these monopoles seems to be
collimated in the vortex direction \cite{VS}.  Finally, it appears that
even the Casimir scaling of higher-representation string-tensions at 
intermediate distance scales can be understood in terms of the finite 
thickness of center vortices \cite{Cas}.

   On these grounds, we are confident that the
vortices identified by our gauge-fixing + projection procedure are
physical objects which are crucial to the confinement mechanism.
But a disquieting question remains, namely: Why does this procedure work?
In what way does the gauge choice \rf{e1}, combined with the
projection \rf{e2}, identify center vortices?  In particular, since
a vortex creation operator (unlike a monopole creation operator)
makes no reference whatever to any special gauge choice or
Higgs field, why do we need to fix to a definite gauge in order
to locate vortices?  These questions become quite urgent when it
is recognized that apparently minor changes in the gauge-fixing condition,
or even, as shown recently by Kov\'{a}cs and Tomboulis \cite{KT}, a small
change in the gauge-fixing \emph{procedure}, can be catastrophic, and
the resulting P-vortices no longer correspond to anything physical.
So what crucial property of the gauge-fixing/projection procedure has
been lost, when the method fails?

   In this article we identify this crucial property of our procedure
as the ``vortex-finding'' (VF) property, by which we mean the following:
Suppose, in any given thermalized lattice, a center vortex is inserted 
``by hand'' via a discontinuous gauge transformation.  The lattice
now contains at least one center vortex in a known location.  Upon 
gauge-fixing and center projection, a set of P-vortices is identified.
\emph{Is the vortex inserted by hand found among this set of P-vortices?}
If so, then the procedure has the vortex-finding property.  This property
seems like a reasonable demand to make of any method which is advertised 
to extract the vortex content of lattice configurations,
and is presumably a necessary condition for its success.

\section{The VF-Property in Adjoint Gauges}
   
   Let us examine why maximal center gauge and, as we will see, 
an infinite class of other gauges, might have this vortex-finding property.
We begin by noting that the maximal center gauge defined by eq.\
\rf{e1} is really the adjoint Landau gauge; i.e.\ it is equivalent to
a Landau gauge-fixing condition on adjoint links
\beq
       R = \sum_x \sum_\m \mbox{Tr}[U_{A\m}(x)] 
             ~~~ \mbox{is a maximum}
\label{e3}
\eeq
where $U_{A\m}(x)$ is the link variable in the adjoint representation.
This motivates considering other gauge-fixing conditions of the
general form
\beq
      {\cal R}[U_A] ~~~ \mbox{is a maximum}
\label{e4}
\eeq
such that the gauge condition
\begin{enumerate}
\item  depends only on the adjoint representation links;
\item  is a complete gauge-fixing of the adjoint link variables;
\item  transforms most links to be close to center elements,
at weak coupling.  
\end{enumerate}
We will refer to these as ``adjoint'' gauges.
   
   Let $U$ denote some thermalized lattice configuration.  A center
vortex is created, on the background $U$, by a discontinuous gauge
transformation (some explicit examples will be given below).  Denote
the resulting configuration as $U'$, whose field strength differs from 
that of $U$ only at the vortex core.  Away from the vortex core, the
corresponding link variables of each configuration in the adjoint 
representation, denoted $U_A$ and $U'_A$, are gauge-equivalent.  This
is because the global discontinuity of the vortex-creating gauge 
transformation, given by a center element, is invisible in the adjoint
representation.  

   The crux of the argument is this: It is assumed that the gauge-fixing
condition \rf{e4} is complete for adjoint links.  If we for the moment
ignore both (i) the Gribov copy problem; and (ii) the region of the
lattice corresponding to the core of the created
vortex, then $U_A$ and $U'_A$ are gauge equivalent, and gauge-fixing
to \rf{e4} should map both $U_A$ and $U'_A$ into
the \emph{same} gauge-fixed configuration $\tU_A$.  Under these
mappings, the link variables $U$ and $U'$ in the fundamental representation
are transformed to configurations $\tU$ and $\tU'$ corresponding to the
same adjoint configuration $\tU_A$.  Since they correspond to the same
SO(3) configuration, the $\tU$ and $\tU'$ lattice 
SU(2) configurations can differ 
only by continuous and/or discontinuous $Z_2$ transformations.  Then,
because a 
continuous gauge transformation, associated with the gauge-fixing, cannot 
undo the discontinuous transformation which created the center vortex, 
the vortex originally inserted in $U'$ appears as a discontinuous $Z_2$
transformation relating $\tU'$ to $\tU$.  

   What has happened here is that the original discontinuous gauge
transformation, which may be quite smooth (up to the discontinuity) 
and extended, has been squeezed
by the gauge-fixing condition to the identity everywhere 
except on a Dirac volume (bounded by the
vortex core), where it has the effect of simply multiplying a certain set 
of links by $-1$.  Upon center projection, $\tU \ra Z$ and $\tU' \ra Z'$, 
and the projected configurations differ by the same discontinuous $Z_2$ 
gauge transformation.  This discontinuity then shows up as an additional 
P-vortex  in $Z'$, not present in $Z$, at the location of the vortex 
inserted by hand.

   This ``vortex-finding property'' goes a long way towards explaining
the success of maximal center gauge in extracting the vortex content of
lattice configurations.  If, in fact, lattice vacuum configurations have 
the form of a product of vortex-creation operators which operate on a 
non-confining background state, then a procedure with the VF-property may 
be reasonably expected to locate these confining vortices.  The above
argument not only explains why maximal center gauge should have the
VF-property, but also suggests that \emph{any} complete gauge-fixing
condition on adjoint links might have the VF-property as well.  However,
there are two ways that the argument we have presented may go
wrong:
\begin{description}
\item{\bf Gribov Copies:~} The argument for vortex-finding is based on 
complete adjoint gauge-fixing; i.e.\ any two gauge-equivalent SO(3) 
configurations
should be mapped to a unique adjoint link configuration.  Unfortunately,
standard methods of implementing eq.\ \rf{e4} in practice, 
such as over-relaxation and simulated annealing, usually wind up in 
local maxima of ${\cal R}[U_A]$, known as Gribov copies, rather than the 
global maximum.  So the argument for the VF-property can fail at this point.
\item{\bf Vortex Cores:~} The configurations $U_A$ and $U'_A$ are only 
gauge-equivalent outside the vortex cores.  Because they are gauge-equivalent
outside this (relatively) small region, we have assumed that
$U_A$ and $U'_A$ will transform to the same gauge-fixed SO(3)
configuration $\tU_A$ outside the core region. This assumption, however, 
could simply be wrong. 
\end{description}

   Because of these caveats, we have no proof of the vortex-finding
property in the adjoint gauges.  This is just as well, since we will
soon discuss cases where the property fails.  Nevertheless, we now
have some idea of why the VF-property \emph{might} hold in maximal
center gauge, and a motivation to test this property numerically, both
in maximal center and in other adjoint gauges, to see if it is
correlated with center dominance.  We should also note that our
argument for the VF-property has no loopholes when (i) the
inserted vortex is thin (one lattice spacing thick), so that $U_A$ and
$U_A'$ are SO(3) gauge-equivalent everywhere on the lattice; and (ii)
there is no Gribov problem.  In that case, according to our argument,
center projection is certain to find the inserted vortex, as will be
confirmed in subsection 3.8 below.

\section{Testing the VF-Property}

   We begin by describing a class of discontinuous SU(2) gauge 
transformations $g_V$ on an $L^3 \times T$ lattice, which create two parallel 
thin vortex surfaces at time $t=T$ that are closed by lattice periodicity.  
These transformations have the usual form 
\beq
      U_\m(\vec{x}) \ra U'_\m(\vec{x}) = 
           g_V(\vec{x}) U_\m(\vec{x}) g_V^{\dg}(\vec{x}+\hat{\m})
\label{e5}
\eeq
except that the gauge transformation has the discontinuity
\beq
        g_V(x,y,z,T+1) = \left\{ \begin{array}{rc}
            - g_V(x,y,z,1) & x_1 \le x \le x_2 \cr
                  g_V(x,y,z,1)     & \mbox{otherwise}
                \end{array} \right.
\label{e6}
\eeq
Its not hard to see that each discontinuous transformation $g_V$ of this form
is equivalent to the mapping
\beq
       U_4(x,y,z,T) \ra - U_4(x,y,z,T) ~~~~
         \mbox{for} ~~ x \in [x_1,x_2]
\label{e7}
\eeq
with all other links unchanged, followed by an ordinary \emph{continuous}
gauge transformation 
\beq
      U_\m(\vec{x}) \ra U'_\m(\vec{x}) = 
           g(\vec{x}) U_\m(\vec{x}) g^{\dg}(\vec{x}+\hat{\m})
\label{r}
\eeq

\begin{figure}[h!]
\centerline{\scalebox{0.5}{\includegraphics{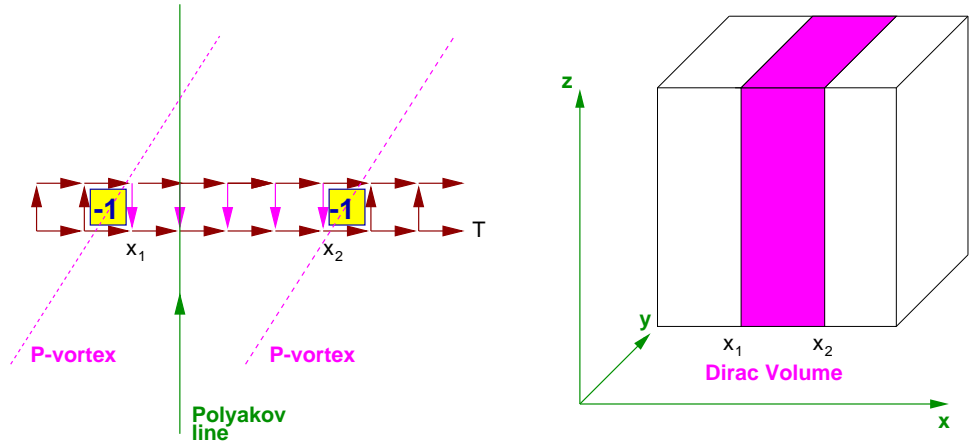}}}
\caption{Left: The effect of the transformation \rf{e7} on 
links at fixed $y,z$ and $t=T,1$.  Downward pointing time-like 
links have been multiplied by $-1$.  Right: Parallel vortex 
surfaces created at constant $x_1,x_2$, $t=T$ bound a Dirac volume, closed in
the $y,z$ directions by lattice periodicity.}
\label{fig1}
\end{figure}

   Suppose that the transformation by $g_V$ is performed on a 
thermalized lattice,
and that the configurations $U$ and $U'$ are gauge-fixed to an adjoint
gauge, and then center-projected.  The corresponding projected configurations
are denoted $Z$ and $Z'$, and we denote the Polyakov lines in these
projected configurations by $P(x,y,z)$ and $P'(x,y,z)$ respectively.
If the gauge-fixing + projection
procedure has the vortex-finding property, then the $Z'$ lattice should
contain P-vortex surfaces on the dual lattice at fixed $t=T$ and fixed $x=x_1-1,x_2$.
The two parallel surfaces bound a
Dirac 3-volume, e.g.\ as indicated in
fig.\ \ref{fig1} (apart from its boundary, the location of the Dirac
volume is gauge-dependent).  Now, if all the \emph{other}
P-vortices were located in identical positions in $Z$ and $Z'$ (i.e.\
if, apart from the inserted vortex, $Z$ and $Z'$ were equivalent up to
a $Z_2$ gauge transformation), then
the vortex-finding property is verified if
\beq
       P'(x,y,z) P(x,y,z) = \left\{ \begin{array}{cc}
                -1 & x \in [x_1,x_2] \cr
                +1 & \mbox{otherwise}
                \end{array} \right.
\label{e8}
\eeq

   At this point, however, we have to face up to the Gribov copy problem.
Center vortices on unprojected lattice configurations are comparatively
thick ($\approx$ 1 fm) objects, and the precise ``middle'' of a vortex
core, which P-vortices are supposed to locate, is a little ambiguous.
In fact, it is found that if we take two gauge copies $U_I$ and
$U_{II}$ of the same
configuration $U$, gauge-fix each to maximal center gauge and then center
project, that the P-vortices in the corresponding projected configurations
$Z_I$ and $Z_{II}$, although correlated, differ somewhat in position.  Thus, 
a P-vortex plaquette in $Z_I$ which is just inside the minimal surface of
a given loop $C$, may lie just outside this minimal surface in configuration
$Z_{II}$ (for a more detailed discussion and related numerical results,
see ref.\ \cite{Jan98}).   
   
  The randomizing effect of small differences in P-vortex
locations in $Z$ and $Z'$ (which correspond to the same ``thick'' vortices in
$U$ and $U'$) will cause 
the vev $\langle P'(x,y,z) P(x,y,z)\rangle$ to differ from $-1$ and $+1$, 
respectively, inside and outside the region $x \in [x_1,x_2]$.  However,
if the vortex-finding property is valid, then eq.\ \rf{e8} should
hold true when this randomizing effect is factored out.
To this end, we generate a set of
thermalized SU(2) lattice configurations and, from each configuration,
we obtain three configurations consisting of
\begin{description}
\item{I.~~} The original configuration, denoted $U_I$.
\item{II.~} The original configuration with an inserted vortex, denoted
$U_{II}$. In most of our simulations, a thin vortex is inserted 
by applying the mapping \rf{e7}, followed by a random (but continuous) 
gauge transformation $g_{II}$ at every site.
\item{III.} A gauge copy of $U_I$, denoted $U_{III}$, obtained
by applying a random continuous gauge transformation $g_{III}$ to $U_I$
at every site.
\end{description}
We then fix each of these configurations to an appropriate adjoint gauge,
center project according eq.\ \rf{e2}, and compute the ratio
\beq
     G(x)=\frac{\sum_{y,z} \langle P_{I}(x,y,z) P_{II}(x,y,z)\rangle}
     {\sum_{y,z} \langle P_{I}(x,y,z) P_{III}(x,y,z)\rangle}
\label{e9}
\eeq
where $P_I,P_{II},P_{III}$ denote Polyakov lines in the corresponding
projected configurations, and
where the denominator has the effect of factoring out the randomizing
(Gribov-copy) effects just mentioned.  Then the gauge-fixing + projection
procedure has the vortex-finding property only if
\beq
       G(x) = \left\{ \begin{array}{rl}
            -1 & x \in [x_1,x_2] \cr
            +1 & \mbox{otherwise} \end{array} \right.
\label{e10}
\eeq

\subsection{Thin Vortex Insertion, Maximal Center Gauge}

   For our first test of the VF-property, we have performed the
Monte Carlo simulation on a $14^3\times 12$ lattice at $\b=2.3$, and
applied a discontinuous gauge transformation (eq.\ \rf{e7} followed
by a random continuous gauge transformation) to insert thin vortices 
as described above. We then fix to maximal center gauge via
over-relaxation, as described in ref.\ \cite{Jan98}, center project
according to eq.\ \rf{e2}, and calculate $G(x)$.    The discontinuity 
in eq.\ \rf{e6} is chosen to lie in the range $x\in [4,10]$.

\begin{figure}[ht!]
\centerline{\scalebox{0.85}{\includegraphics{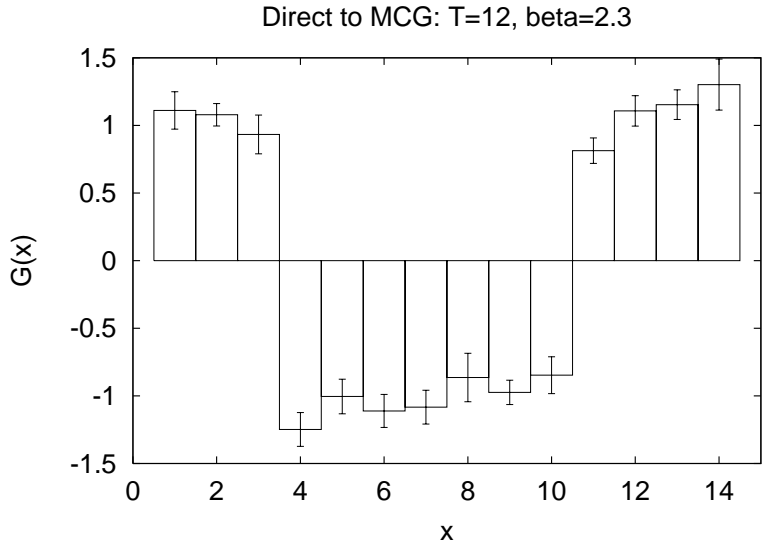}}}
\caption{Graph of $G(x)$ for configurations
 with thin inserted vortices ($14^3\times 12$ lattice, 380 lattices, 
$\beta=2.3$).  The gauge discontinuity \rf{e7} is located in the 3-volume at 
 $x\in[4,10]$, $t=T$.}
\label{fig2}
\end{figure}

   The result for $G(x)$ is shown in fig.\ \ref{fig2}.  The criterion
for the vortex-finding property, eq.\ \rf{e10}, appears to be nicely
confirmed in this case.  Given the successes of our procedure, summarized
in the Introduction, in extracting the vortex content of vacuum 
configurations, the existence of a vortex-finding property was perhaps to
be expected.  We find that in this case, the existence of Gribov copies 
in our gauge-fixing procedure does not destroy the VF-property.

   Instead of inserting vortices via the transformation \rf{e7},
followed by a random gauge transformation \rf{r}, we have also
considered vortex insertion by gauge transformations which are smooth in
the t-direction up to
the discontinuity at $t=T+1$, taking the form
\beq
        g_V(x,y,z,t) = \left\{ \begin{array}{cl}
            \exp\Bigl[i\pi (t-1) \s_3/T \Bigr] & \mbox{for} ~~ 
                                           x\in [x_1,x_2] \cr
                 1     & \mbox{otherwise}
                \end{array} \right.
\eeq
and performed simulations for the same parameters as before
($\b=2.3$, $14^3\times 12$ lattice, $[x_1,x_2]=[4,10]$).  As in the
previous case, the results are consistent with the vortex-finding
property in eq.\ \rf{e10}.

   Apart from our tests here, based on the values of $G(x)$, we     
would also like to mention some relevant results reported
recently by Montero in ref.\ \cite{Montero}. Montero, building on the 
work of ref.\ \cite{Tony}, constructs
classical SU(3) center vortex solutions on a periodic lattice.  The
stability of the solution is due to the use of twisted boundary
conditions, with two dimensions of the periodic lattice chosen
much smaller than the two other dimensions.  The lattice is fixed
to maximal center gauge and then center-projected.  It is found that
P-vortex plaquettes accurately locate the middle of the classical
vortex.  This is very interesting independent evidence for the existence of 
the vortex-finding property of maximal center gauge (in SU(3) 
gauge theory, in this case).

\subsection{Thin Vortex Insertion, Kov\'{a}cs-Tomboulis Procedure}

   The argument for the vortex-finding property of adjoint gauges,
as presented in section 2, explicitly neglects the problem of Gribov copies,
and in fact the possibility of gauge-fixing to a unique configuration was 
an essential step in the argument.  Conversely, it follows
that the existence of Gribov copies in the gauge fixing
procedure might destroy the
vortex-finding property, and, as a consequence, the ability of the
projection procedure to locate center vortices in thermalized lattice
configurations.  This appears to be what happens in a modification
of our gauge-fixing procedure considered recently by Kov\'{a}cs and Tomboulis 
\cite{KT} (and also noted in ref.\ \cite{dFE}.)
   
   Kov\'{a}cs and Tomboulis suggest fixing first to the (usual) lattice
Landau gauge, before fixing to maximal center gauge via over-relaxation.
Of course, if the over-relaxation procedure transformed each configuration
to a unique, global maximum of eq. \rf{e1} (i.e.\ if there were
no Gribov copies), then a preliminary fixing to lattice Landau gauge
could not possibly make a difference to the result.  However, the
over-relaxation procedure only finds a \emph{local} maximum of 
\rf{e1}.  Thus the starting point can make a difference, at least in
principle.\footnote{There is not much difference, however, in $R$. Making a
set of gauge copies of a given configuration, and gauge-fixing
each copy to maximal center gauge either directly, or with a Landau-gauge
starting point, we find that the variation in $R$ of eq.\ \rf{e1}
among different copies far exceeds the average difference in $R$
between the two procedures; in fact, our own results for this average 
difference are not yet statistically significant.}  Kov\'{a}cs and 
Tomboulis find that with a Landau gauge starting
point, center dominance is completely lost in the projected configurations,
which appear to have no asymptotic string tension.  This fact does not
really affect the status of
P-vortices, identified by the usual method, as locators of physical objects.
That status is well established by the correlation of P-vortices with 
gauge-invariant observables, and by their scaling properties.
In the modified procedure, however, it would appear that
some important property, essential for extracting the center vortex
content of vacuum configurations, has been lost.  From everything that
has been said until now, the obvious candidate for this ``essential
property'' is the vortex-finding property.

\begin{figure}[ht!]
\centerline{\scalebox{0.85}{\includegraphics{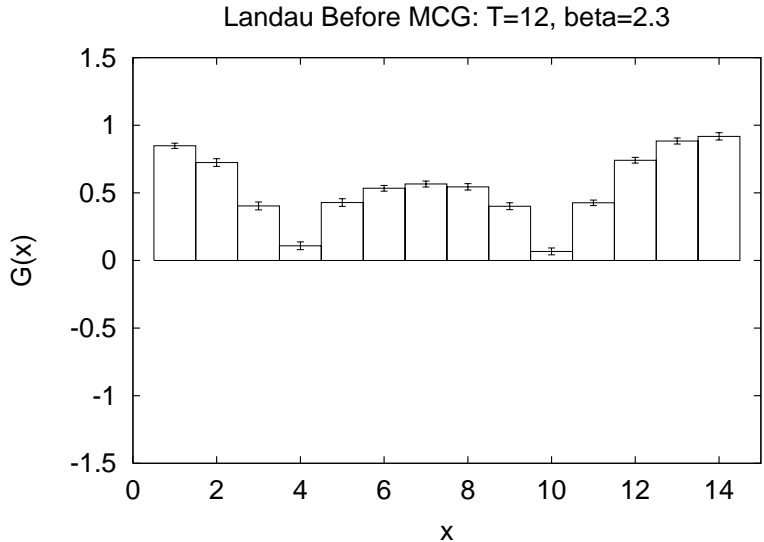}}}
\caption{Graph of $G(x)$ for configurations
 with a thin inserted vortex. Configurations are
 first fixed to the Landau gauge, and only then to maximal center gauge
 ($14^3\times12$ lattice, 260 lattices, $\beta=2.3$).
 The Dirac volume is the same as in fig.\ \ref{fig2}.}
\label{fig3}
\end{figure}

   In fig.\ \ref{fig3} we display our results for the observable $G(x)$,
in which each configuration $U_I,U_{II},U_{III}$ is first fixed to
Landau gauge, before fixing to maximal center gauge via over-relaxation.
The contrast between Figs.\ \ref{fig2} and \ref{fig3} is quite striking;
the Kov\'{a}cs-Tomboulis procedure is clearly inconsistent with eq.\ \rf{e10},
and seems to have completely lost the vortex-finding property.  Only
at the boundary of the Dirac volume, where there is a strong local
field strength, does $G(x)$ show some effect (although $G(x)>0$ even there).
But in the middle of the Dirac volume, $G(x)$ is comparable to its value
outside the volume.  It seems likely that if the interval $[x_1,x_2]$ were
large enough, the middle of the Dirac volume would be indistinguishable
from the outside region.  Since this modified procedure cannot even identify a 
thin center vortex inserted 
into the lattice by hand, there is no reason to expect it to locate the
fuzzy, much more diffuse center vortices generated by the gauge theory
dynamics.

\subsection{Thin Vortex Insertion, Asymmetric Adjoint Gauge}

    The argument for the vortex-finding property in section 2 did
not single out the maximal center gauge in particular; it is possible
that other adjoint gauge choices might work just as well.
Let us therefore introduce the ``Asymmetric Adjoint Gauge''
\beq
       R = \sum_x \sum_\m c_\m \Bigl| \mbox{Tr}[U_\m(x)] \Bigr|^2
             ~~~ \mbox{is a maximum}
\label{e11}
\eeq
where $c_\m$ is some set of four positive numbers.  For an exploratory
run at $\b=2.3$, again on a $14^3\times 12$ lattice, we chose a  
set of values
\beq
     \{c_1,c_2,c_3,c_4\} = \{1.0,1.5,0.75,1.0\}
\label{e12}
\eeq
and fixed to the gauge \rf{e1} by over-relaxation (without prior
Landau gauge-fixing).  

\begin{figure}[ht!]
\centerline{\scalebox{0.85}{\includegraphics{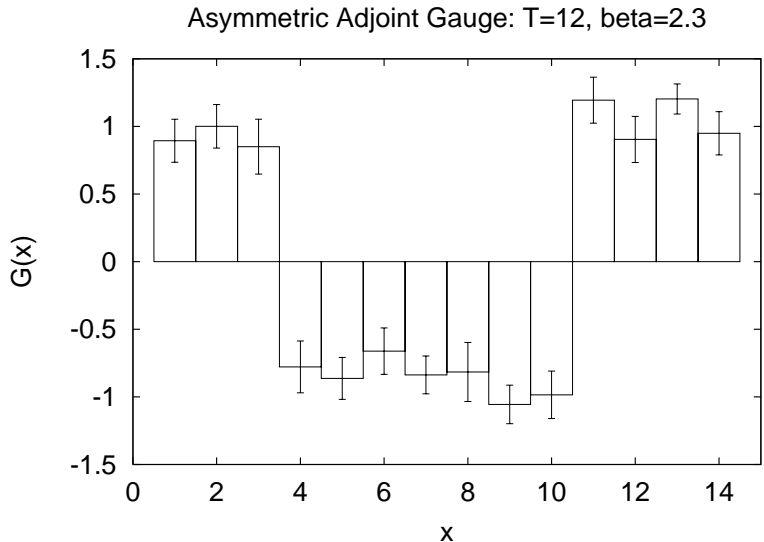}}}
\caption{Graph of $G(x)$ for configurations
 with a thin inserted vortex. Configurations are
 fixed to the asymmetric adjoint gauge with 
 $c_\mu=\{1,1.5,0.75,1\}$ ($14^3\times 12$ lattice, 
 380 lattices, $\beta=2.3$).
 The Dirac volume is the same as in fig.\ \ref{fig2}.}
\label{fig4}
\end{figure}

\begin{figure}[ht!]
\centerline{\scalebox{0.5}{\includegraphics{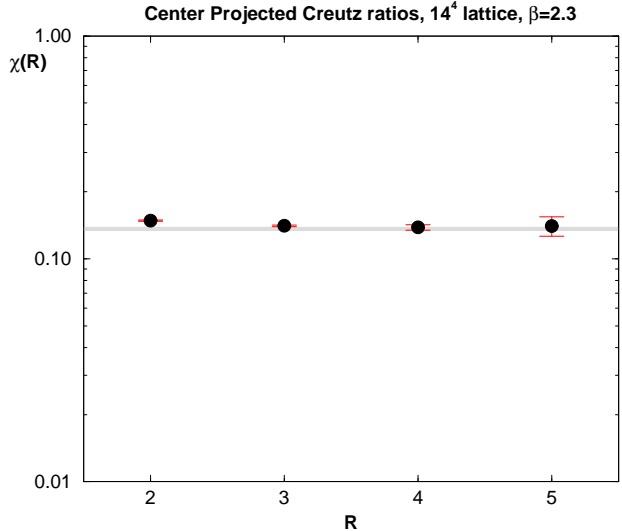}}}
\caption{Center dominance for the gauge with 
 $c_\mu=\{1,1.5,0.75,1\}$. ($14^4$ lattice, $\beta=2.3$.)
 Solid line shows the value of asymptotic string tension
 quoted in ref.\ \cite{Bali}.}
\label{fig5}
\end{figure}

   The result for $G(x)$ in this new gauge is shown in fig.\ \ref{fig4}.
Once again, the vortex-finding property is satisfied within
errorbars.  The next question is: Do the corresponding center-projected
configurations display center dominance (i.e.\ do they have the same
string tension as the unprojected configuration)?  In a run at
$\b=2.3$ on a $14^4$ lattice, in the asymmetric adjoint gauge \rf{e11},
\rf{e12}, the projected Creutz ratios do, in fact, agree quite well with the 
asymptotic string tension extracted from gauge-invariant Wilson loops,
reported in ref.\ \cite{Bali}.  The results are shown in fig.\ \ref{fig5}.

\subsection{The Modulus Landau and Adjoint Coulomb Gauges}

   It was certainly not obvious that the preliminary Landau gauge-fixing, 
suggested by Kov\'{a}cs and Tomboulis, followed by maximal center gauge-fixing
via over-relaxation, would destroy the VF-property (and, as
a consequence, center dominance) in the projected configurations.
It is also a little surprising that certain choices of adjoint gauge, 
which would naively seem just as good as maximal center gauge, appear to have  
similar problems. An example is a ``modulus'' version of the usual
lattice Landau gauge
\beq
       R = \sum_x \sum_\m \Bigl| \mbox{Tr}[U_\m(x)] \Bigr|
             ~~~ \mbox{is a maximum}
\label{e12a}
\eeq
which is also an adjoint gauge, as defined by the conditions set out
in section 2.  We believe, on the basis of the argument in section 2, 
that if $R$ could be fixed to a unique global
maximum then this gauge would also have the VF-property, and the projected
configurations would exhibit center dominance.  However, as with
maximal center gauge, the only known gauge-fixing techniques are
simulated annealing and over-relaxation, and these are both plagued
with Gribov copies.  So the only way of testing the vortex-finding
property, and center dominance, is numerically.

\begin{figure}[h!]
\centerline{\scalebox{0.85}{\includegraphics{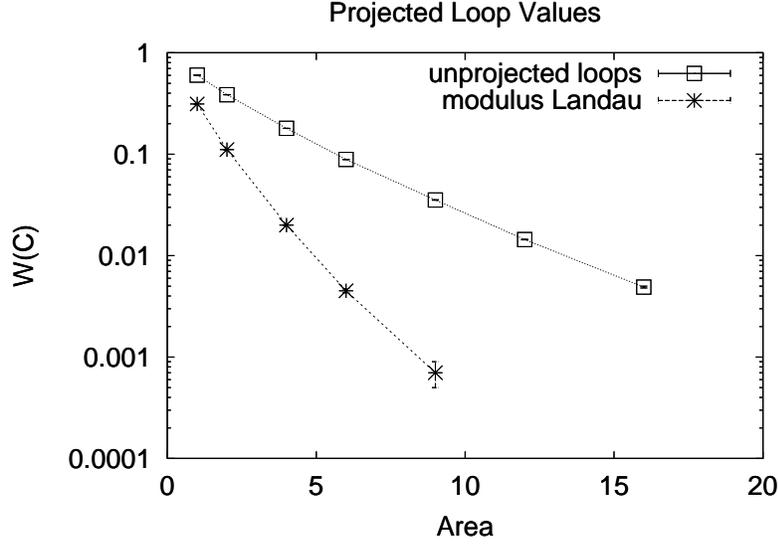}}}
\caption{Center-projected Wilson loops vs.\ Area in modulus Landau gauge,
for square $R\times R$ and rectangular $R\times (R+1)$ loops, at $\b=2.3$ on a
$12^4$ lattice.  Unprojected Wilson loop values are also shown
for comparison.}
\label{ablan1}
\end{figure}

   We have used over-relaxation at $\b=2.3$ to fix to the modulus
Landau gauge.  In this case, the falloff with area of the projected
Wilson loops appears to be much faster than that of the unprojected
loops, as seen in fig.\ \ref{ablan1}; center dominance is clearly lost.

\begin{figure}[h!]
\centerline{\scalebox{0.85}{\includegraphics{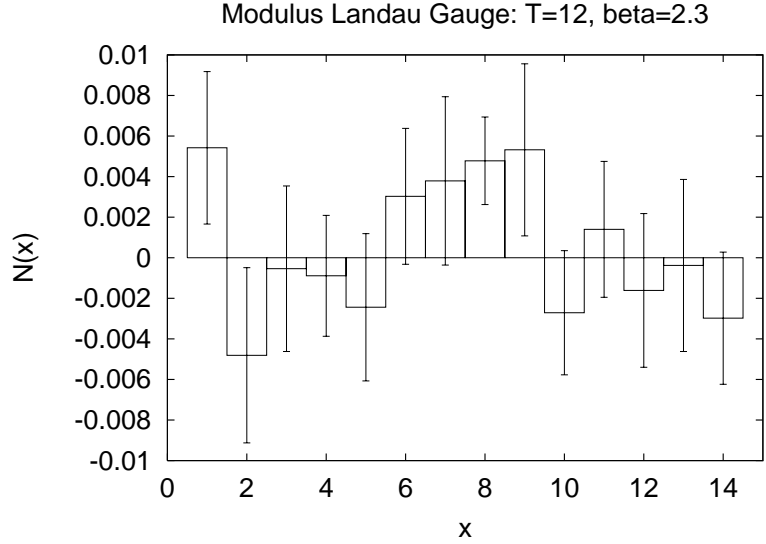}}}
\caption{Numerator $N(x)$ (eq. \rf{e12c}) of the ratio $G(x)$, for 
modulus Landau gauge ($14^3\times 12$ lattice, 380 lattices,
$\beta=2.3$).  The Dirac volume is the same as in fig.\ \ref{fig2}.}
\label{ablan}
\end{figure}

    The corresponding values for $G(x)$ in modulus Landau gauge have very 
large errorbars, and
this is simply because both the numerator and denominator in eq.\ \rf{e9}
have values which are, within errorbars, consistent with zero.  Results
for the product in the numerator
\beq
     N(x) ={1\over L^2} \sum_{y,z} \langle P_{I}(x,y,z) P_{II}(x,y,z)\rangle
\label{e12c}
\eeq
are shown in fig.\ \ref{ablan}, where $L$ ($=14$) is the lattice length in the
$y,z$-directions.  It is clear that there is in this case
a total loss of the vortex-finding property.
In contrast to the Kov\'{a}cs-Tomboulis case,
in which $G(x)$ was positive irrespective of linking to the inserted
vortex, in this case $G(x)$ seems to be essentially $0/0$, irrespective of
linking number.  But just as in the Kov\'{a}cs-Tomboulis case, the failure of
center dominance in the projected configurations is associated with a 
corresponding loss of the vortex-finding property.

\begin{figure}[h!]
\centerline{\scalebox{0.85}{\includegraphics{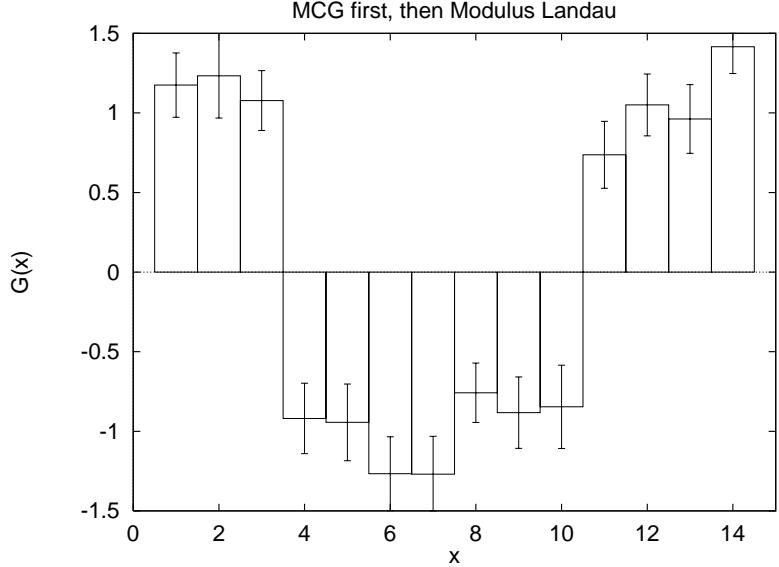}}}
\caption{Graph of $G(x)$ for configurations
 with thin inserted vortices ($14^3\times 12$ lattice, 110 lattices, 
$\beta=2.3$) in modulus Landau gauge.  In this case, the lattice
is first fixed to maximal center gauge, before fixing via over-relaxation to
modulus Landau gauge.}
\label{abnot}
\end{figure}

   Since the modulus Landau gauge should be a perfectly good adjoint
gauge, the failure of the vortex-finding property in this case 
must be attributed to the Gribov copy problem.  This impression is strengthened
when one compares the rms value of $a_0$, where
\beq
         U_\m = a_0 I + i \vec{a} \cdot \vec{\s}
\eeq
in both the modulus Landau and maximal center gauges.  We find that
$a_0^{rms} = 0.76$ in modulus Landau gauge, vs.\ $a_0^{rms} = 0.86$
in maximal center gauge, at $\b=2.3$, and that is a surprisingly large 
discrepancy, in view of the fact that both gauges are, in some sense,
trying to bring links close to center elements.  This suggests trying
out a variation of the Kovacs-Tomboulis procedure: Perhaps 
modulus Landau gauge-fixing could be improved, if we first fix to maximal 
center gauge, and \emph{then} fix to modulus Landau gauge via over-relaxation.

   It turns out that a preliminary gauge-fixing to maximal center
gauge restores both center dominance and the vortex-finding property.
For center dominance, we find that the Creutz ratios, at $\b=2.3$, 
are clustered near $\chi[R,R] \approx 0.14$, quite close to the asymptotic
string tension obtained on unprojected lattices.  The result for $G(x)$ 
is shown in fig.\ \ref{abnot}; it is similar to what we previously found
for maximal center gauge and the asymmetric Landau gauge. 

   Once again, the presence or absence of the vortex-finding property
is associated with the presence or absence of center dominance in the
projected configurations.

\begin{figure}[h!]
\centerline{\scalebox{0.85}{\includegraphics{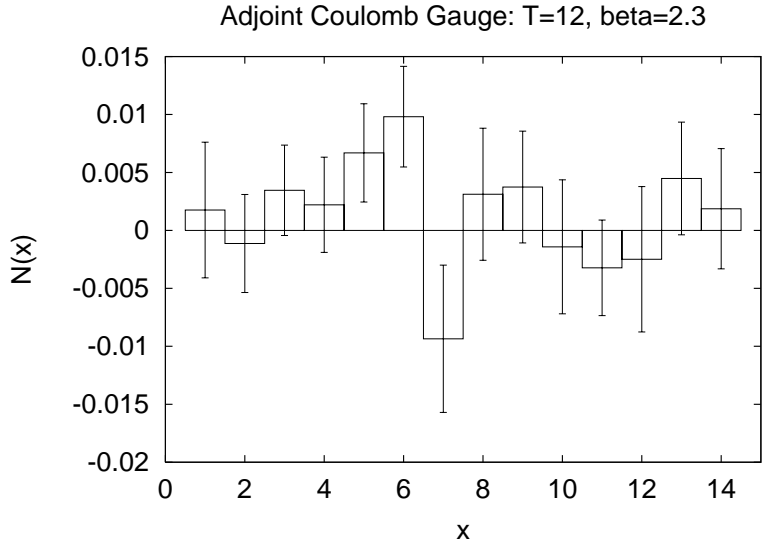}}}
\caption{Numerator $N(x)$ (eq. \rf{e12c}) of the ratio $G(x)$, for 
adjoint Coulomb gauge ($14^3\times 12$ lattice, 180 lattices, 
$\beta=2.3$).  The Dirac volume is the same as in fig.\ \ref{fig2}.}
\label{adjcoul}
\end{figure}

   In section 3.3 we studied the asymmetric adjoint gauge, with
a moderate variation of $c_\m$ between $0.75$ and $1.5$.  It is also
of interest to look at limiting cases of this gauge, where the ratios
of some of the $c_\m$'s tend to zero or infinity.
A particular example
we have investigated is the adjoint Coulomb gauge, in which
\beq
     \{c_1,c_2,c_3,c_4\} = \{1.0,1.0,1.0,0.0\}
\label{e12d}
\eeq
Center dominance is lost in this gauge also, although the disagreement between
full and projected string tensions is not quite as bad as in 
modulus Landau gauge without preconditioning.  Creutz ratios in the 
projected configurations
at $\b=2.3$ cluster around $\chi_{proj}(R,R) \approx 0.18$, whereas
the full asymptotic string tension is around $\s a^2 = 0.135$.  

   As in modulus Landau gauge, the loss of center dominance in
adjoint Coulomb gauge is accompanied by a breakdown of the vortex-finding
property, as  seen from a plot of
the numerator \rf{e12c} in fig.\ \ref{adjcoul}.  The qualitative difference
between adjoint Coulomb gauge, and the asymmetric adjoint gauge studied
in section 3.3, may be connected with the fact that adjoint
Coulomb gauge is not really an adjoint gauge, as defined in section 2.
Our third criterion for adjoint gauges is that link variables should
be brought close to center elements.  What we find instead, at
$\b=2.3$, is that while the rms value $a_0^{rms}$ for spacelike
links is $0.89$, the corresponding rms value for timelike links
is only $0.50$ (and this is identical to the rms values of the other
three components $a_k^{rms}$.  Thus, while spacelike links do indeed 
fluctuate near center elements, the timelike links do not, 
and the ``close-to-center'' criterion is violated.  
This criterion for adjoint gauges will be further discussed in section 3.6.

\subsection{Thicker Inserted Vortex, Maximal Center Gauge}

   All of the inserted vortices so far are thin vortices; the vortex
core has a thickness of one lattice spacing.  On the other hand, the vortices
generated dynamically by the gauge theory seem to be rather thick objects, 
with core widths on the order of one fermi.  It is therefore of interest
to see if an inserted vortex with a somewhat thickened core can also
be identified on the projected lattice.

  In our previous examples, the $-1$ discontinuity in the
vortex-creating transformation $g_V$ begins and ends abruptly
at $x=x_1$ and $x=x_2$.  To create a vortex core several lattice
units thick in the x-direction, we simply make the transition from
the $-1$ discontinuity to $+1$ continuity more gradual.  This is
done by replacing the mapping in eq.\ \rf{e7} by  
\beq
       U_4(x,y,z,T) \ra  U_4(x,y,z,T) \exp[i\a(x)\s_3] 
\label{e13}
\eeq
where $\a(x)$ interpolates smoothly from $\a(x)=0$ outside the
Dirac volume, to $\a(x)=\pi$ inside the Dirac volume.  

   We make the choice for $\a(x)$ shown in Table \ref{table1}
on an $18^3\times 8$ lattice, followed by a random gauge transformation, to
obtain configurations $U_{II}$.  

\begin{table}[h!]
\centerline{
\begin{tabular}{|c||c|c|c|c|c|c|c|c|c|c|c|c|c|c|c|c|c|c|} \hline\hline
  x&1&2&3& 4&  5&  6&  7&  8&  9& 10& 11& 12& 13& 14&15&16&17&18 \\ \hline 
$ \a$ &0&0&0& $\q$ & ${2\pi\over 5}$ & ${3\pi\over 5}$ & 
${4\pi\over 5}$ & $\pi$ & $\pi$ & $\pi$ & $\pi$ &$ {4\pi\over 5}$ &
${3\pi\over 5}$ & ${2\pi\over 5}$ & $\q$ & 0&0&0 \\ \hline
\end{tabular} }
\caption{Thick vortex core: $\a(x)$ in eq.\ \rf{e13}.}
\label{table1}
\end{table}

\begin{figure}[ht!]
\centerline{\scalebox{0.85}{\includegraphics{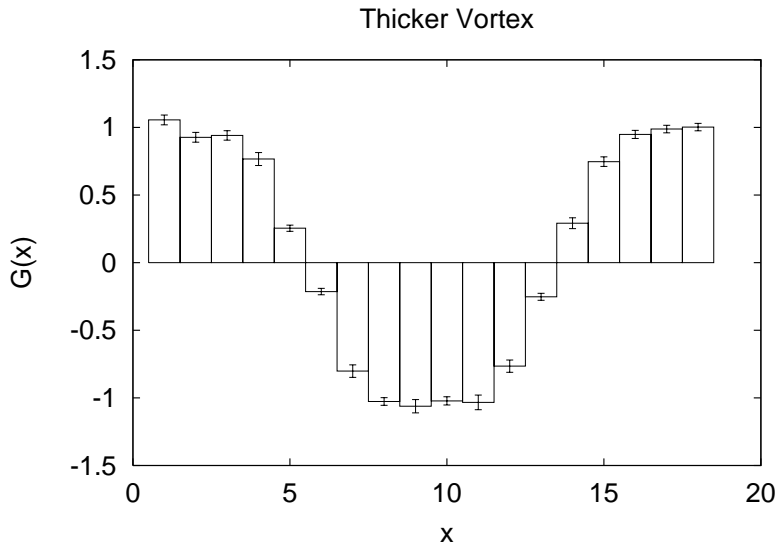}}}
\caption{$G(x)$ computed for thicker
 vortices on an $18^3\times 8$ lattice, 380 lattices, $\beta=2.3$.}
\label{fig7}
\end{figure}

\ni Maximal center gauge-fixing and
center projection are carried out, and $G(x)$ is calculated,
with results shown in fig.\ \ref{fig7} (obtained at $\b=2.3$).   
Once again the vortex-finding property is quite evident, with
$G(x)$ interpolating smoothly from $+1$ to $-1$ across the
vortex core.

\subsection{The ``Close-to-Center'' Criterion}

   All of the gauges we have considered have the property of bringing
most link variables close to the $\pm I_2$ center variables.  This
was listed as the third criterion of an adjoint gauge in section 2,
and we should now explain the rationale behind this criterion.  

   The picture which is implied by the numerical studies [1$-$11]
is that thermalized SU(2) lattice configurations have the form
\beq
       U = G_V \circ U_{NC}
\eeq
where $G_V$ is an operator creating center vortices responsible for
confinement, while $U_{NC}$ is a non-confining lattice background.
The operator $G_V$ is a ``smoothed'' discontinuous gauge transformation;
i.e.\ it is has the form of a discontinuous gauge transformation, 
away from vortex cores.\footnote{If $G_V$ had this form everywhere, then it
would produce thin vortices of very high action; instead, the regions
of high field strength are smoothed out, and thick vortices are created.} 
In an adjoint gauge which has the vortex-finding property, center projection
should locate the approximate position of vortices created by $G_V$.

   There is no guarantee, however, that gauge-fixing + center projection
will not also produce, in addition to the P-vortices associated with $G_V$, a
lot of other extraneous P-vortices with no physical content, nor is
there any obvious reason that these extraneous P-vortices should 
be independent of the choice of adjoint gauge.  This is the motivation 
for introducing, in the definition of adjoint gauges, a ``close-to-center'' 
requirement.

   In maximal center gauge (or any other adjoint gauge), not all links are 
close to center elements, even at large $\b$; some links lying in the middle of
center vortices must deviate very substantially from $\pm I_2$.
This deviation from the center elements
is necessary, in order that a projected plaquette can equal $-1$, while the
corresponding gauge-invariant plaquette on the unprojected lattice is 
close to $+1$ at weak coupling (cf. \cite{Tub0}).  The aim of the 
``close-to-center'' requirement is that links should deviate substantially
from center elements only where they \emph{must} do so, i.e.\ in the
cores of vortices created by $G_V$; elsewhere the gauge-fixing condition
should force links to fluctuate in the vicinity of $\pm I_2$.  This
criterion is, of course, most obvious in maximal center gauge.

   Let MC denote maximal center gauge, and AG denote another adjoint
gauge which (i) has the VF-property, and (ii) has most links close to
center elements, except in the middle of center vortices.  We can then
argue that, away from vortex interiors,
the center-projected configurations $Z_{MC}$ and $Z_{AG}$ will be
$Z_2$ gauge-equivalent.  Let $U'_{MC},U'_{AG}$ represent a thermalized
configuration $U$ fixed to the MC and AG gauges, 
respectively.  We can write
\bea
          U'_{MC} &=& Z_{MC} U''_{MC} ~~~ \mbox{where} ~~~
                  U''_{MC} = \mbox{signTr}[U'_{MC}] U'_{MC}
\non \\
          U'_{AG} &=& Z_{AG} U''_{AG} ~~~~ \mbox{where} ~~~
                  U''_{AG} ~ = \mbox{signTr}[U'_{AG}] U'_{AG}
\eea
where the product and sign trace operations are of course performed 
link by link.\footnote{The $U''_{MC}$ configuration, incidentally, is the 
lattice studied by de Forcrand and D'Elia \cite{dFE}, and it is found to
have neither confining nor chiral-symmetry breaking properties.}
Because $U'_{MC}$ and $U'_{AG}$ are gauge-equivalent, we have
\bea
       U'_{AG} &=& (zg) \circ U'_{MC}
\non \\
               &=& (z \circ Z_{MC}) (g \circ U''_{MC})
\eea
where $z$ is a $Z_2$ gauge transformation, and $\mbox{Tr}[g] \ge 0$.
Away from vortex interiors, most links in $U'_{MC}$ and $U'_{AG}$ are
close to $\pm I_2$ at large $\b$, which implies that $g \circ U''_{MC}
\approx \pm I_2$ are also close to center elements.  Combining this
with the fact that $U''_{MC}\approx + I_2$ and $\mbox{Tr}[g]>0$, we
deduce that $g(x) \approx g(x+\mu)$ in this region, and therefore $g
\circ U''_{MC} \approx +I_2$.  We can then identify
\bea
        U''_{AG} &=& g \circ U''_{MC}
\non \\
        Z_{AG} &=& z \circ Z_{MC}
\eea
According to this argument, $Z_{AG}$ and $Z_{MC}$ are $Z_2$
gauge-equivalent, away from the vortex interior.

   Inside a vortex core, $U''_{MC}$ is not necessarily close to the
identity, so $Z_{AG}$ and $Z_{MC}$ are not necessarily
gauge-equivalent.  This implies that there will in general be some
variation in P-vortex location among different adjoint gauges (and
among Gribov copies in the same adjoint gauge), associated with the
finite width of the center vortex core.  But apart from this local
variation, the vortices found by two adjoint gauges should be
basically the same, providing that both gauges have the vortex-finding
property, and providing that the links in both gauges are close to
center elements everywhere except where they \emph{must} strongly
deviate, i.e. in the core of vortices created by $G_V$.  How well this
``close-to-center'' requirement is actually fulfilled, in different
adjoint gauges with the vortex-finding property, is a topic which has
not yet been studied in any detail.

\subsection{Speculations about Cooling and Smoothing}

   If maximal center gauge + center projection is applied to 
cooled configurations, then the projected string tension is drastically
reduced after only a few cooling steps (similar results are found for 
RG-smoothed configurations) \cite{Jan98}.  But it was also found 
in ref.\ \cite{Jan98} that the cores of thick center vortices, whose 
approximate location is identified before cooling,
also expand considerably after only a few cooling steps, as measured
by the one-vortex to zero-vortex loop ratio $W_1(C)/W_0(C)$. 

   Center vortices in thermalized lattices have been found to be rather
thick, ``fuzzy'' objects, which percolate throughout the lattice.  It 
appears that, after a few cooling steps, the vortex cores overlap so much
that there is virtually no region of the lattice which is \emph{not} part of
a vortex core.  In view of the second caveat (``Vortex Cores'')
in section 2, it is then not so surprising that the maximal center gauge +
center projection procedure is unable to extract the locations of these
very fat, highly overlapping objects (although the method seems to have no 
trouble finding thin, inserted vortices after a few cooling steps).

\subsection{Laplacian Center Gauge}

   According to our argument in section 2, an adjoint gauge without the Gribov
copy problem is \emph{guaranteed} to have the vortex-finding property,
at least for thin inserted vortices where vortex width is only one lattice
spacing.  Such a gauge was invented recently by Alexandrou,
D'Elia, and de Forcrand \cite{dFE1}; it is known as the ``Laplacian
Center Gauge.''

   To fix to Laplacian center gauge, one finds the
eigenvectors $\vec{u}(x),\vec{v}(x)$ corresponding to the two lowest
eigenvalues of of the covariant lattice Laplacian  
\beq
      \D_{xy}^{ab} = 2D \d_{xy} \d^{ab} - \sum_{\pm \m}
                        U_{A,\pm \m}^{ab}(x) \d_{x\pm \m,y}
\eeq
where
\beq
        U^{ab}_{A,\m} = \oh \mbox{Tr}[U_\m \s^a U_\m^\dg \s^b]
\eeq
are matrix elements of the link variables in the adjoint representation.
The gauge transformation which takes the lattice configuration into
Laplacian center gauge is the transformation which rotates $\vec{u}(x)$
to lie in the 3-direction, and $\vec{v}(x)$ to lie in the 1-3 plane, at
every site $x$.  This procedure is free of Gribov copies, and it was
found in the second reference of \cite{dFE1} that there is agreement
between the asymptotic string tensions on the full and projected
lattices, at least at the coupling $\b=2.4$ used there.  Although
Laplacian center gauge does not involve directly maximizing a
functional (as in eq.\ \rf{e4}), it does satisfy all three conditions
below \rf{e4}, and thus qualifies as an adjoint gauge.

%\begin{figure}[ht!]
%\centerline{\scalebox{0.85}{\includegraphics{lcg.eps}}}
%\caption{Graph of $G(x)$ for configurations
% with a thin inserted vortex. Configurations are
% fixed to the Laplacian center gauge on a
% $14^3\times 12$ lattice at $\beta=2.3$.
% The Dirac volume is the same as in fig.\ \ref{fig2}.}
%\label{lcg}
%\end{figure}

   We have checked the VF-property for Laplacian center gauge as
before; i.e.\ for a thin inserted vortex on the same lattice as
in fig.\ \ref{fig2}, at $\b=2.3$.  The somewhat startling result is that
the condition
\beq
       G(x) = \left\{ \begin{array}{rl}
            -1 & x \in [4,10]  \cr     
            +1 & \mbox{otherwise} \end{array} \right.
\label{exact}
\eeq
is satisfied by the numerical data \emph{exactly}.
In other words, there are
no errorbars at all; every configuration gives the same result
for $G(x)$.  Fixing to Landau gauge, prior to Laplacian
center gauge, makes no difference to this result.

   In fact, the absence of errorbars on $G(x)$ in Laplacian center
gauge is simply a confirmation of the reasoning in sections 2 and 3.
As noted above, the VF-property is guaranteed in this gauge, at least
for thin vortices.  We have also argued, in section 3, that any
deviation from eq.\ \rf{e8} must be due to small variations in
P-vortex location, from one Gribov copy to another, and these variations
are responsible for the statistical fluctuations in $G(x)$.  Eliminating
Gribov copies eliminates in turn these statistical fluctuations, and
the VF-property for Laplacian center gauge is seen in the most
compelling way.

   Laplacian center gauge is not quite as ``close to
center'' as maximal center gauge.  We find, for example, that
the rms value of $a_0 = \mbox{Tr}[U]/2$ at $\b=2.3$ is approximately 
$a_0^{rms}=0.86$ in maximal center gauge, but only $a_0^{rms}=0.82$ in 
Laplacian center gauge. There
are also more P-vortex plaquettes in the projected Laplacian configurations;
ref.\ \cite{dFE1} reports an excess in P-vortex plaquettes 
of about $11$\%, at $\b=2.4$, as compared to maximal center gauge. 
The greater number of P-vortex plaquettes found in Laplacian center gauge
does not affect the asympotic string tension of projected configurations, 
and can probably be attributed to one (or both) of two 
sources: (i) a roughening of the P-vortex surface; 
(ii)  some extraneous ``small'' P-vortices, unconnected to the 
single large vortex responsible for confinement, which is found 
\cite{bertle} to percolate through the entire lattice.
Either of these effects could modify the projected Creutz ratios at
short distances, while preserving the asymptotic string tension. 
This may explain the slightly delayed approach of projected Creutz ratios
to their asymptotic value,  relative to previous
results in maximal center gauge.

   Alexandrou et al.\ in ref.\ \cite{dFE1} also suggest an
alternative method for identifying vortices via Laplacian center gauge, which 
is based on locating certain gauge-fixing ambiguities, rather than
center projection.  We have not yet investigated this method; it would 
be interesting to know if our argument for the vortex-finding property in 
adjoint gauges can also be applied in this alternative framework.

\section{Conclusions}

   Gauge-fixing has had a bad reputation, in connection with studies
of the confinement mechanism.  One is justifiably suspicious of
any calculation which depends in an essential way on some special
gauge choice, particularly if the physical motivation of that special
gauge choice is unclear.  In this article we hope to have dispelled
some of that suspicion, at least in connection with maximal center 
and related gauges. Center vortices are created by discontinuous gauge
transformations, which of course make no reference to any particular gauge 
condition.  We have argued above that in maximal center gauge $-$ 
and in an infinite class of other adjoint gauges $-$ such
discontinuous transformations are squeezed to the identity everywhere 
except on Dirac volumes, whose locations (together with those of 
the associated vortices) are then revealed upon center projection.  This 
is the ``vortex-finding property'' which motivates the use of
adjoint gauges, and it explains
how adjoint gauge-fixing, combined with center projection, can extract
the vortex content of thermalized lattice configurations.  

   While we think it likely that every adjoint gauge, as defined
in section 2, has the VF-property, this is certainly not true of every 
gauge-fixing \emph{procedure}.  The argument for the VF-property 
assumed a complete and unique adjoint gauge-fixing, but unfortunately
the standard over-relaxation and
simulated annealing methods are plagued by Gribov copies.  This is a
large loophole in the argument for vortex-finding, and in certain cases, 
e.g.\ the Kov\'{a}cs-Tomboulis procedure, the Gribov copy problem is 
apparently severe enough to destroy the VF-property.  

    It is not clear, at present, why the
Gribov problem destroys the VF-property only in some 
gauge-fixing procedures, but not in others.  What does seem
clear, however, is that the vortex-finding property, and the center dominance
of projected configurations, go hand-in-hand.  
Maximal center gauge, asymmetric adjoint gauge, modulus Landau gauge
with maximal center preconditioning, and 
Laplacian center gauge all have the VF-property, and all exhibit
center dominance.  Conversely, in the (i) Kov\'{a}cs-Tomboulis procedure, 
(ii) modulus Landau gauge without preconditioning, 
and (iii) adjoint Coulomb gauge, the VF-property is lost, and in 
none of these cases is there center dominance in the projected configurations.

   We conclude with a sort of tautology: \emph{To find center vortices,
one must use a procedure with the vortex-finding property}.  If the
gauge-fixing + projection procedure 
doesn't have the VF-property, or if that property is 
destroyed by some modification (e.g.\ by Landau gauge preconditioning),
then center vortices are not correctly identified on thermalized lattices, 
and center dominance
in the projected configuration is lost.  This fact does not call
into question the physical relevance of P-vortices found by our usual
procedure (which \emph{has} the vortex-finding property); that 
relevance is well established by the strong correlation that exists
between these objects and gauge-invariant observables.  Ideally, potential
problems due to Gribov copies could be avoided altogether by
a gauge-fixing procedure which fixes to a unique adjoint link configuration.
Laplacian center gauge \cite{dFE1} appears to be an example of just such a
procedure.

\vspace{32pt}

\ni {\Large \bf Acknowledgements}

\bigskip

  We thank Phillipe de Forcrand for discussions.
J.G.\ is happy to acknowledge the hospitality of the high-energy theory
group at the Niels Bohr Institute, where much of this
work was carried out.  

   Our research is supported in part by Fonds zur F\"orderung der
Wissenschaftlichen Forschung P11387-PHY (M.F.), the U.S. Department of 
Energy under Grant No.\ DE-FG03-92ER40711 (J.G.), and the Slovak Grant 
Agency for Science, Grant No. 2/4111/97 (\v{S}. O.).

\end{document}